\documentclass[11pt]{article}
\usepackage{amsmath,amssymb,color,graphics,epsfig,cite}

\usepackage{amsfonts}

\newcommand{\hoch}[1]{$\, ^{#1}$}

\newcommand{\be}{\begin{equation}}
\newcommand{\ee}{\end{equation}}
\newcommand{\bea}{\setlength\arraycolsep{2pt} \begin{eqnarray}}
\newcommand{\eea}{\end{eqnarray}}
\newcommand{\nn}{\nonumber}

\def\ft#1#2{{\textstyle{\frac{\scriptstyle #1}{\scriptstyle #2} } }}
\def\fft#1#2{{\frac{#1}{#2}}}

\def\0{{\sst{(0)}}}
\def\1{{\sst{(1)}}}
\def\2{{\sst{(2)}}}
\def\3{{\sst{(3)}}}
\def\4{{\sst{(4)}}}
\def\5{{\sst{(5)}}}
\def\6{{\sst{(6)}}}
\def\7{{\sst{(7)}}}
\def\8{{\sst{(8)}}}
\def\sst#1{{\scriptscriptstyle #1}}
\def\oneone{\rlap 1\mkern4mu{\rm l}}

\begin{document}

\begin{center}
{\Large {\bf Dyonic Black Strings and the Charge Lattice\\ in Salam-Sezgin Model}}

\vspace{20pt}

{\large Liang Ma\hoch{1}, Yi Pang\hoch{1} and H. L\"u\hoch{1,2}}

\vspace{10pt}

{\it \hoch{1}Center for Joint Quantum Studies and Department of Physics,\\
School of Science, Tianjin University, Tianjin 300350, China }

\bigskip

{\it \hoch{2}Joint School of National University of Singapore and Tianjin University,\\
International Campus of Tianjin University, Binhai New City, Fuzhou 350207, China}

\vspace{40pt}

\underline{ABSTRACT}
\end{center}

We obtain a class of dyonic black string solutions in 6D Salam-Sezgin model. We then calculate various thermodynamic quantities associated with this solution. Interestingly, for the thermodynamic quantities to be well defined, the temperature is bounded from above. However, the mass and entropy can still grow without any upper bound, reaching infinity at the maximal temperature. The quantization condition obeyed by various charges is also analyzed. In particular, we find that the Dirac quantization condition selects one particular sign choice for the magnetic string charges.

\vfill{\footnotesize liangma@tju.edu.cn\ \ \ pangyi1@tju.edu.cn\ \ \ mrhonglu@gmail.com
 }


\thispagestyle{empty}
\pagebreak



\section{Introduction}

Salam-Sezgin model is the simplest six-dimensional (6D) $N=(1,0)$ supergravity model in which the R-symmetry group $U(1)_R$ is gauged \cite{Salam:1984cj}. This gauging of the $U(1)_R$ symmetry generates a positive definite potential given by the exponential function of the dilaton. Consequently, the vacua are not maximally symmetric spacetime, but instead Minkowski$_4\times S^2$. This feature makes the model appealing in cosmology and phenomenology \cite{Maeda:1984gq,Maeda:1985es, Gibbons:1986xp,Aghababaie:2003ar,Aghababaie:2003wz}. The original Salam-Sezgin model, despite being anomalous on its own, can be extended to anomaly-free models by enlarging extra matter content \cite{Randjbar-Daemi:1985tdc, Avramis:2005qt} and adopting the Green-Schwarz mechanism. A drawback of these anomaly free extensions is that due to the mixing between tree-level and one-loop effects and the inclusion of anomaly-induced higher curvature terms, the fully supersymmetric field equations are only known partially \cite{Pang:2020rir}.  This makes the study of the original anomalous model still worthwhile since it can be viewed as the tree-level approximation in the long-wavelength limit.

Some attempts of embedding the original Salam-Sezgin model in string/M-theory have been made.  For instance, an non-compact embedding of the theory in M-theory was obtained in \cite{Cvetic:2003xr}, by consistently truncating the resulting $N=(1,1)$ theory to $N=(1,0)$.
However, there is no convincing mechanism to remove the additional modes outside the model. On the other hand, this does provide an uninspiring resolution of the anomalous problem of the original $N=(1,0)$ theory. Interestingly, the bosonic sector of the theory can also be embedded into a variant $N=(1,1)$ gauged supergravity via generalized Kaluza-Klein reduction of M-theory on K3$\times \mathbb R$ \cite{Kerimo:2003am,Kerimo:2004md}. Furthermore, it turns out to be consistent to perform the full $S^2$ reduction keeping all the $SO(3)$ gauge fields \cite{Gibbons:2003gp}, giving rise to Einstein-Yang-Mills theory without a cosmological constant.

It is well-known that the ungauged model ($g=0$)  admits asymptotically flat dyonic string solutions \cite{Duff:1993ye,Duff:1996cf} which play an important role in deriving the criteria for 6D ungauged $N=(1,0)$  supergravities to acquire a UV completion \cite{Seiberg:2011dr, Monnier:2017oqd}.  By analogy understanding properties of dyonic string solutions in Salam-Sezgin model may be used as an independent tool to decipher whether the original Salam-Sezgin model could admit a UV completion regardless of string theory embedding. We recall that a $\ft14$-BPS (supersymmetry) dyonic string solution in Salam-Sezgin model was obtained in \cite{Gueven:2003uw} which showed that the near horizon limit of the solution is AdS$_3\times$ squashed $S^3$ while at asymptotic infinity, it approaches a cone over Minkowski$_2\times$ squashed $S^3$ as opposed to Minkowski$_4\times S^2$.

In this paper, we show that the supersymmetric dyonic string solution in Salam-Sezgin model can be heated up to acquire a non-zero temperature. We then calculate various thermodynamic quantities associated with this solution. Interestingly, for the thermodynamic quantities to be well defined, the temperature is bounded from above by a $T_{\rm max}$ depending on the charges and the gauging coupling $g$. However, the mass and entropy can still grow without any upper bound, reaching infinity at the maximal temperature. We also analyze the Dirac quantization condition obeyed by various charges. This is an nontrivial exercise because some of the charge parameters in the original dyonic string solutions of \cite{Gueven:2003uw} are subject to an algebraic constraint that is inconsistent with the Dirac quantization. We find that the genuine dyonic strings reside in a new parameter region beyond the original one.

The paper is organized as follows. In section \ref{sec:bpsdyon1}, we review the BPS dyonic string of \cite{Gueven:2003uw}. We study its global structure and show that the geodesic is incomplete using the original radial coordinate and we give the new radial coordinate. In section \ref{sec:black}, we construct the non-extremal extension of the BPS dyon and analyse the black hole thermodynamics. In section 4, we treat the original BPS solution on its own and derive its first law. In section \ref{sec:lattice}, we study the charge lattice of the dyonic strings, based on the Dirac quantization condition. We show that the charge lattice of the original BPS dyon is vacuous and the nontrivial charge lattice arises from the new parameter regions. We conclude our paper in section \ref{sec:conclusion}.

\section{The BPS dyon and its global structure}
\label{sec:bpsdyon1}

The original Salam-Sezgin does admit a Lagrangian formulation and the bosonic sector is \cite{Salam:1984cj}
\bea
e^{-1}\mathcal{L}_6&=& R-\frac{1}{4}\left(\partial\varphi\right)^2-\frac{1}{12}e^\varphi H_\3^2-\frac{1}{4}e^{\frac{1}{2}\varphi}F_\2^2-8g^2e^{-\frac{1}{2}\varphi},\nn\\
F_{\2}&=&dA_{\1},\qquad H_\3=dB_\2+\frac{1}{2}F_{\2}\wedge A_{\1}\,,
\label{SS}
\eea
where $g$ is the $U(1)_R$ gauge coupling. (Note that on the left-hand side, $e=\sqrt{-\det(g_{\mu\nu})}$.) The complete set of equations of motion derived from the Lagrangian \eqref{SS} is give by
\bea
0&=&R_{\mu\nu}-\frac{1}{2}g_{\mu\nu}\,e^{-1} \mathcal{L}_6-
\frac{1}{4}\left(\partial_\mu\varphi\right)\left(\partial_\nu\varphi\right)
-\frac{1}{4}e^\varphi H_{\mu\alpha\beta}H_{\nu}^{\ \alpha\beta}-\frac{1}{2}e^{\frac{1}{2}\varphi}F_{\mu\alpha}F_{\nu}^{\ \alpha},\cr
0 &=&-\frac{1}{2}d {\star d\varphi} -\frac{1}{2}e^\varphi {\star {H}_{(3)}}\wedge {H}_{(3)}
-\frac{1}{4}e^{\frac{1}{2}\varphi}{\star F_{(2)}}\wedge F_{(2)}+4g^2e^{-\frac{1}{2}\varphi}{\star \oneone}\,,\cr
0&=&d\left(e^{\frac{1}{2}\varphi}{\star F_{(2)}}+\frac{1}{2}e^\varphi {\star {H}_{(3)}}\wedge A_{(1)}\right)-\frac{1}{2}e^\varphi {\star {H}_{(3)}}\wedge F_{(2)}\,,\cr
0&=&d\left(-e^\varphi {\star {H}_{(3)}}\right).\label{6D EOM form fields}
\eea
Turning on the $H_\3$ field strength gives rise to dyonic string solutions. The second-order partial differential equations of motion can be in general difficult to solve. The equations reduce to differential equations if one imposes spherically symmetric and static conditions. However, the spherical symmetry can be broken to $SU(2)\times U(1)$ isometry if we turn on the magnetic dipole charge associated with $F_\2$. Supersymmetry reduces the second-order differential equations to the first order by imposing the Killing spinor equations
\bea
&&(\nabla_\mu - {\rm i} g A_\mu + \ft1{48} e^{\fft12\phi} H_{\alpha\beta\gamma} \Gamma^{\alpha\beta\gamma} \Gamma_\mu)\epsilon=0\,,\cr
&&(\Gamma^\mu \partial_\mu \phi - \ft16 e^{\fft12\phi} H_{\mu\nu\rho} \Gamma^{\mu\nu\rho}) \epsilon=0\,,\qquad (e^{\fft14\phi} F_{\mu\nu}\Gamma^{\mu\nu} - 8{\rm i} g e^{-\fft14 \phi})\epsilon=0\,.
\eea
It was found that the theory admits a dyonic BPS string solution \cite{Gueven:2003uw}
\bea
ds^2&=&H_P^{-\frac{1}{2}}H_Q^{-\frac{1}{2}}\left(-dt^2+dx^2\right)
+\frac{k^2P}{4g^2r^6}H_P^{-\frac{5}{2}}H_Q^{\frac{1}{2}}dr^2
+\frac{k}{4g}H_P^{-\frac{1}{2}}H_Q^{\frac{1}{2}}
\left(\sigma^2_1+\sigma^2_2+\frac{4gP}{k}\sigma^2_3\right),\cr
A_{\1}&=&-k\cos\theta d\varphi,\quad B_{\2}=\left(H_Q^{-1}+c\right)dt\wedge dx-P\cos\theta d\varphi\wedge d\psi\,,\cr
H_Q&=&Q_0+\frac{Q}{r^2},\qquad H_P=P_0+\frac{P}{r^2},\qquad \varphi=\log \fft{H_Q}{H_P}\,,
\label{bpsdyon}
\eea
where some of the parameters are subject to an algebraic constraint
\be
4gP=k(1-2gk)\,.\label{gkpcons}
\ee
The $(\sigma_1,\sigma_2,\sigma_3)$ are three $SU(2)$-left invariant 1-forms
\be
\sigma_1=\cos\psi d\theta+\sin\psi\sin\theta d\varphi\,,\qquad \sigma_2=-\sin\psi d\theta+\cos\psi\sin\theta d\varphi\,,\qquad \sigma_3=d\psi+\cos\theta d\varphi\,.
\ee
The volumes of unit $S^2$ ($ds_2^2=\sigma_1^2 + \sigma_2^2$) and $S^3$ ($ds_3^2=\fft14(\sigma_1^2 + \sigma_2^2 + \sigma_3^2))$ are
\be
\omega_2 \equiv \int \sigma_1\wedge \sigma_2=4\pi\,,\qquad
\omega_3 \equiv \fft18 \int \sigma_1\wedge\sigma_2\wedge \sigma_3 = 2\pi^2\,.
\ee

The BPS dyonic string solution \eqref{bpsdyon} carries both electric and magnetic string charges, given by
\be
Q_e = \fft{1}{16\pi} \int e^{\varphi} {\star H_\3} = \pi Q\,,\qquad
Q_m = \fft{1}{16\pi} \int dB_\2 = \pi P\,.\label{QeQm}
\ee
The solution also carries magnetic dipole charge associated with $F_\2$, given by
\be
Q_D=\fft{1}{16\pi} \int F_\2\wedge \sigma_3=\fft{1}{4} \int F_\2=\pi k\,.
\label{QD}
\ee
The convention of the overall factor for the dipole charge can be derived from the Wald formalism discussed later. The parameters $(P_0,Q_0)$ are related to the speed of light and the scalar modulus at asymptotic $r\rightarrow \infty$.

The global structure of the BPS dyon was discussed in \cite{Gueven:2003uw}. In particular, $r\rightarrow 0$ gives rise to the near-horizon geometry of direct product of AdS$_3$ and a {\it squashed} $S^3$. In fact, the theory also admits AdS$_3 \times S^3$, but it is not supersymmetric. It is worth pointing out that the AdS$_3 \times S^3$ vacuum is supersymmetric in the variant $N=(1,1)$ gauged supergravity \cite{Kerimo:2003am,Kerimo:2004md}. The $r\rightarrow \infty$ limit, on the other hand, remains murky. One puzzling feature is that $g_{rr}\sim 1/r^6$ at large $r$ vanishes, indicating that the proper radial distance is not infinity for large $r$. This is related to another puzzling fact that
although the constraint \eqref{gkpcons} have a smooth limit to $g\rightarrow 0$ and $k\rightarrow 0$, with fixed $P$, the resulting dyonic solution of ungauged supergravity does not have the usual form.  To understand the latter puzzle, the metric was cast into the following form in \cite{Gueven:2003uw}:
\be
ds^2 = (H_P H_Q)^{-\fft12} (-dt^2 + dx^2) + (H_P H_Q)^{\fft12}\Big(\fft{4\xi^2 dr^2}{\Xi^3} +
\fft{r^2}{\Xi}(\sigma_3^2 + \xi (\sigma_1^2 + \sigma_2^2))\Big)\,,
\ee
where
\be
\xi=\fft{k}{4gP} = (1-2g k)^{-1}\,,\qquad \Xi=1 + \fft{P_0}{P} r^2\,.
\ee
In the limit of $g\rightarrow 0$ and $k\rightarrow 0$, one has $\xi\rightarrow 1$. It was stated in \cite{Gueven:2003uw}, that the solution becomes the standard dyonic string solution in this limit provided that $\Xi=1$, which requires $P_0=0$. This seems to suggest that there are two different types of dyonic strings, one with $P_0=0$ and one without. To see this, we should redefine the radio coordinate
\be
\fft{r^2}{\Xi} = \rho^2\,,\label{rhodef}
\ee
for which the solution \eqref{bpsdyon} becomes
\bea
ds^2 &=& (H_P H_Q)^{-\fft12} (-dt^2 + dx^2) + (H_P H_Q)^{\fft12}\Big(4\xi^2 d\rho^2 +
\rho^2 (\sigma_3^2 + \xi (\sigma_1^2 + \sigma_2^2))\Big)\,,\nn\\
A_{\1}&=&-k\cos\theta d\varphi,\quad B_{\2}=(H_Q^{-1}+c)dt\wedge dx-P\cos\theta d\varphi\wedge d\psi\,,\cr
H_P&=& \fft{P}{\rho^2}\,,\qquad H_Q=Q_0 - \fft{P_0 Q}{P} + \fft{Q}{\rho^2}\,.\label{bpsdyonrho}
\eea
Note that in this parametrization, we should set $P_0=0$, since $P_0$ appears only in $H_Q$, where it can be absorbed into $Q_0$.  Furthermore, we can add an additional pp-wave component in the string direction, namely
\be
-dt^2 + dx^2 \rightarrow -H_K^{-1} dt^2 + H_K (dx + H_K^{-1} dx)^2\,,\qquad H_K = 1 + \fft{K}{\rho^2}\,.\label{ppwave}
\ee
Finally we note that the scalar field is given by
\be
e^{\varphi} = \fft{Q_0}{P} \rho^2 + \fft{Q}{P}\ge \fft{Q}{P}\,,\label{scalarsol}
\ee
for all the $\rho\ge 0$ region.

\section{Blackening the BPS dyonic string}
\label{sec:black}

The blackening of dyonic string of ungauged supergravity, corresponding to setting both $g$ and $k$ to zero in \eqref{bpsdyonrho}, is straightforward, using the standard procedure of constructing the non-extremal black $p$-branes \cite{Duff:1996hp,Cvetic:1996gq}. We can also follow the same procedure to obtain the non-extremal dyonic string with non-vanishing $g$. In the original $r$ coordinate of \eqref{bpsdyon}, we find that the solution is given by
\bea
ds^2&=&H_P^{-\frac{1}{2}}H_Q^{-\frac{1}{2}}\left(-fdt^2+dx^2\right)
+\frac{k^2P}{4g^2r^6}H_P^{-\frac{5}{2}}H_Q^{\frac{1}{2}}\frac{dr^2}{f}
+\frac{k}{4g}H_P^{-\frac{1}{2}}H_Q^{\frac{1}{2}}
\left(\sigma^2_1+\sigma^2_2+\frac{4gP}{k}\sigma^2_3\right),\cr
A_{\1}&=&-k\cos\theta d\varphi,\quad B_{\2}=\left(\sqrt{1+\frac{\mu  \left(P Q_0-P_0 Q\right)}{P Q}}H_Q^{-1}+c\right)dt\wedge dx-P\cos\theta d\varphi\wedge d\psi\,,\cr
H_Q&=&Q_0+\frac{Q}{r^2},\qquad H_P=P_0+\frac{P}{r^2},\qquad \varphi=\log \fft{H_Q}{H_P}
\,,\qquad f=1-\frac{\mu P_0}{P}-\frac{\mu}{r^2}\,.\label{non-BPS-r}
\eea
In the $\rho$ coordinate defined by \eqref{rhodef}, we find that the solution becomes
\bea
ds^2 &=& (H_P H_Q)^{-\fft12} (-fdt^2 + dx^2) + (H_P H_Q)^{\fft12}\Big(\frac{4\xi^2 d\rho^2}{f} +\rho^2 (\sigma_3^2 + \xi (\sigma_1^2 + \sigma_2^2))\Big)\,,\cr
A_{\1}&=&-k\cos\theta d\varphi,\quad B_{\2}=\left(\sqrt{1+\frac{\mu  \left(P Q_0-P_0 Q\right)}{P Q}}H_Q^{-1}+c\right)dt\wedge dx-P\cos\theta d\varphi\wedge d\psi\,,\cr
H_P&=& \fft{P}{\rho^2}\,,\qquad H_Q=Q_0 - \fft{P_0 Q}{P} + \fft{Q}{\rho^2}\,,\qquad f=1-\frac{\mu}{\rho^2}\,,\label{non-BPS-rho}
\eea
where we can set $P_0=0$ without loss of generality. Note that we can set integration constant $(\mu,Q)$ to zero freely, but not the $(P,k)$ parameters that satisfy the constraint \eqref{gkpcons}.  The solution can be viewed as an non-extremal electrically-charged string in the magnetic string-dipole background. Note also that we can perform Lorentz boost in the $(t,x)$ direction and the pp-wave \eqref{ppwave} arises as the consequence of infinite boost under the vanishing limit of the blackening parameter $\mu$.

It is therefore natural to treat $(\mu,Q)$ as thermodynamic variable and the remaining parameters as thermodynamic constants. We have thus two independent thermodynamic variables. We find that all the thermodynamical variables can be calculated in standard way, given by
\bea
M&=&\pi \Big(\mu + \fft{Q}{Q_0}\Big)\,,\qquad T=\fft{g \sqrt{P\mu}}{k\pi \sqrt{Q + \mu Q_0}}\,,\qquad S=\fft{\sqrt{k^2\mu (Q + \mu Q_0)}\pi^2}{g\sqrt{P}}\,,\nn\\
Q_e &=&\pi \sqrt{Q(Q + \mu Q_0)}\,,\qquad
\Phi_e=\fft{\sqrt{Q}}{Q_0\sqrt{Q + \mu Q_0}}\,.
\eea
(Note that mass and entropy are density quantities per unit string length of $x$.) They satisfy the thermodynamic first law
\be
\delta M = T \delta S + \Phi_e \delta Q_e\,.
\ee
We can express the mass and entropy as functions of $(Q_e,T)$, given by
\bea
M&=&\fft{Q_e}{Q_0 }\left(1-\fft{T^2}{T_{\rm max}^2}\right)^{-\frac{1}{2}},\qquad
\Phi_e=\frac{1}{ Q_0}\sqrt{1-\fft{T^2}{T_{\rm max}^2}}\,,\cr
S &=& \fft{Q_e T}{Q_0 T_{\rm max}^2}\left(1-\fft{T^2}{T_{\rm max}^2}\right)^{-\frac{1}{2}} = \fft{M}{T_{\rm max}^2} T\,,\qquad T_{\rm max} = \fft{1}{\pi}\sqrt{\fft{g^2P}{k^2Q_0}}\,.
\eea
We therefore see that the temperature is bounded, namely
\be
0\le T\le T_{\rm max}\,.
\ee
The lower bound corresponds to the BPS limit with zero temperature.  In the extremal $\mu\rightarrow 0$ limit, the solution reduces to the BPS solution discussed earlier, where both temperature and entropy vanish. At the lower temperature, the entropy is proportional to the temperature.

In the temperature upper bound, both entropy and mass diverge. The specific heat for fixed $Q_e$ is always positive, given by
\be
C_{Q_e} = \fft{1}{T} \fft{\partial S}{\partial T}\Big|_{Q_e} =\fft{Q_0^2M^4}{T_{\rm max}^4 Q_e^2\, S}>0\,.
\ee
It is peculiar that the chemical potential $\Phi_e$ depends only on the temperature, but not the electric charge. We therefore do not have charge capacitance defined in fixed chemical potential.

We can also express the entropy as a function of mass and the electric charge, namely
\be
S=T_{\rm max}^{-1}\sqrt{M^2 - \fft{Q_e^2}{Q_0^2}}\,,\qquad
T= T_{\rm max}\sqrt{1 - \fft{Q_e^2}{Q_0^2M^2}}\,,\qquad \Phi_e=
\fft{Q_e}{M Q_0^2}\,.\label{smq}
\ee
Related to the fact that $\Phi_e$ is a function of $T$ only, the thermodynamic geometry associated with the Hessian metric of the entropy $S(M,Q)$ is one dimensional only.
It follows from \eqref{smq} that the mass and electric charge satisfy the inequality
\be
M\ge \fft{Q_e}{Q_0}\,,
\ee
which saturates in the extremal limit. It is also worth commenting that the Smarr relation is given by
\be
M = T S + \Phi_e Q_e\,.
\ee
This implies that the Gibbs free energy vanishes identically. The Helmholtz free energy is non-negative and hence there is no Hawking-Page type of phase transition.

\section{The first law of the original BPS solution}
\label{sec:floforiginal}

The original BPS dyonic solution \eqref{bpsdyon} constructed in \cite{Gueven:2003uw} has an advantage that we can set both electric and magnetic string charges to zero, leaving only the dipole charge associated with the parameter $k$. It is straightforward to set $Q=0$, but it is more subtle to set $P=0$. To do so, we can first re-scale the coordinates and charge $Q$, namely
\be
r\rightarrow P^\fft14 r\,,\qquad Q\rightarrow \sqrt{P} Q\,,\qquad \psi \rightarrow z P\,.
\ee
We can now set the parameter $P=0$ smoothly, giving rise to $k=1/(2g)$ and
\be
ds^2=H_Q^{-\fft12} (-dt^2 + dx^2) + H_Q^{\fft12} \Big(\fft{dr^2}{16g^4 r^6} + dz^2\Big) +
\fft{1}{8g^2} H_Q^{\fft12} (\sigma_1^2 + \sigma_2^2)\,.
\ee
(Setting $P=0$ in the solution with $\rho$ coordinate is trickier.) We can further set $H_Q=1$ (i.e. $Q_0=1, Q=0$), and obtain the Mink$_4\times S^2$ solution
\be
ds_6^2 = -dt^2 + dx^2 + dy^2 + dz^2 + \fft{1}{8g^2} (\sigma_1^2 + \sigma_2^2)\,,\qquad
F_\2 = \pm \fft{1}{2g} \sigma_1\wedge\sigma_2\,.
\ee
(The coordinate $y$ arises from the redefinition of $r$.)  Thus the original solution \eqref{bpsdyon} can be viewed as dyonic string on the Mink$_4\times S^2$ background. It is thus of interest to study the first law of the original solution with $r$ restricted as a real number running from 0 to infinity.

We adopt the Wald formalism \cite{Wald:1993nt} and we find that the Noether charge and surface terms in the corresponding Wald formalism are
\bea
\mathbf{Q}&=&-{\star d\xi}-e^\varphi {\star H_{\3}}\wedge i_\xi B_{\2}-\left(\frac{1}{2}e^\varphi {\star H_{\3}}\wedge A_{\1}+e^{\frac{1}{2}\varphi}
{\star F_{\2}}\right)\left(i_\xi A_{\1}\right),\cr
\mathbf{\Theta}^\mu_G&=&g^{\alpha\beta}\delta\Gamma^\mu_{\alpha\beta}-
g^{\mu\nu}\delta\Gamma^\alpha_{\alpha\nu}\,,\qquad
\mathbf{\Theta}_B=e^\varphi {\star H_{\3}}\wedge\delta B_{\2}\,,\cr
\mathbf{\Theta}_A&=&-e^{\frac{1}{2}\varphi} {\star F_{\2}}\wedge\delta A_{\1}-\frac{1}{2}e^\varphi {\star H_{\3}}\wedge A_{\1}\wedge\delta A_{\1}\,,\qquad
\mathbf{\Theta}_\varphi = \frac{1}{2} {\star d\varphi}\delta\varphi\,.
\eea
Since both $A_{\1}$ and $B_{\2}$ involve magnetic charges, we need to define a 1-form field $\Psi_{\1}$ and a 2-form field $\Psi_{\2}$ by the EOMs \eqref{6D EOM form fields},
\bea
d\left(-e^\varphi {\star H_{\3}}\right)=0,\qquad &\Rightarrow& \qquad d\Psi_{\1}=i_\xi\left(e^\varphi {\star H_{\3}}\right),\\
d\left(e^{\frac{1}{2}\varphi} {\star F_{(2)}}+e^\varphi {\star H_{(3)}}\wedge A_{(1)}\right)=0,\quad &\Rightarrow& \quad d\Psi_{\2}=i_\xi\left(-e^{\frac{1}{2}\varphi} {\star F_{(2)}}-e^\varphi {\star H_{(3)}}\wedge A_{(1)}\right).\nn
\eea
They are
\be
\Psi_{\1}=-\frac{P}{P_0 \left(P_0 r^2+P\right)}dx\,,\qquad  \Psi_{\2}=\frac{k P}{P_0 \left(P_0 r^2+P\right)}dx\wedge d\psi\,.\label{psi1psi2}
\ee
We may also treat the parameter $g$ in the scalar potential also as a thermodynamic variable, which leads to a 4-form $\chi_{(4)}$, given by
\bea
d\chi_{\4}=-i_\xi\frac{\delta \mathbf{L}}{\delta g^2}\delta g^2,\quad \Rightarrow\quad
\chi_{\4}=\frac{k^2 r^2 }{2g^2 \left(P_0 r^2+P\right)}\delta g^2\sin\theta dx\wedge d\theta\wedge d\varphi\wedge d\psi\,.\label{chi4}
\eea
With these, we can give the improved infinitesimal Hamiltonian density 4-form a la \cite{Lu:2013ura, Ma:2022nwq}
\bea
&&\delta {\cal H}=\delta\mathbf{Q}-i_\xi\mathbf{\Theta}-\chi_{(4)}+d\left(\Psi_{\1}\wedge\delta B_{\2}\right)+d\left(\Psi_{\2}\wedge\delta A_{\1}\right)\cr
=&&\left(\frac{1}{4 g P_0}\delta k-\frac{k}{4 g^2 P_0}\delta g-c\delta Q
\right)\sin\theta dx\wedge d\theta\wedge d\varphi\wedge d\psi\,.
\label{B field H}
\eea
It is clear that 4-form $\delta {\cal H}$ is closed in that $d\delta {\cal H}=0$. In fact, the result is radially independent. Furthermore, it is also closed with respect the variation operator $\delta$, provided that the constraint \eqref{gkpcons} is satisfied. This leads to the definition of mass. If we choose the gauge $c=-\frac{1}{Q_0}$, the electric part of $B_{\2}$ field $\Phi_{e}\delta Q_{e}$ will be present on the horizon $r=0$, but not asymptotic infinity. This quantity can be interpreted as $\delta M$ at $r\rightarrow\infty$, with
\bea
M=\pi\left(\frac{k}{4 g P_0}+\frac{Q}{Q_0}\right).
\eea
The electric/magnetic string charges and the dipole charges are all independent of the radial coordinates and were obtained in \eqref{QeQm} and \eqref{QD} respectively. (Note that the Wald formalism makes it necessary to integrate also over the $\sigma_3$ cycle for computing the dipole charge as in \eqref{QD}.) Their corresponding thermodynamic potentials are given by
\bea
\Phi_{m}&=&\Psi_x|_{r\rightarrow\infty}-\Psi_x|_{r=0}=\frac{1}{P_0}, \qquad \Phi_{e}=B_{tx}|_{r\rightarrow\infty}-B_{tx}|_{r=0}=\frac{1}{Q_0}\,,\nn\\
\Phi_D &=& \Psi_{\psi x}|_{r\rightarrow\infty} - \Psi_{\psi x}|_{r=0} = \fft{k}{P_0}\,.\label{Phis}
\eea
We find that the first law (at zero temperature) and Smarr relation are
\bea
\delta M&=&\Phi_{e}\delta Q_{e}+\Phi_{m}\delta Q_{m}+\Phi_{D}\delta Q_{D},\cr
M&=&\Phi_{e}Q_{e}+\Phi_{m}Q_{m}+\ft{1}{2}\Phi_{D} Q_{D}.
\eea
This first law is very different from the earlier results in $\rho$ coordinate where $(P,k)$ were treated as thermodynamic constants. To understand the above relation in the $\rho$ coordinate, we note that in the Wald formalism, the closure of $d{\delta \cal H}=0$ is independent of the choice of coordinate, and it implies that $\delta {\cal H}|_1 = \delta {\cal H}|_2$, where the subscripts denote the two boundaries.
It follows from \eqref{rhodef} that
\be
r \in [0,\infty) \qquad\leftrightarrow\qquad \rho \in [0,\sqrt{\ft{P}{P_0}}]\,.
\ee
In the $\rho$ coordinates, the potential fields \eqref{psi1psi2} are given by (up to some pure gauge):
\be
\Psi_{\1}=\fft{\rho^2}{P} dx\,,\qquad  \Psi_{\2}=-\fft{k \rho^2}{P} dx\wedge d\psi\,,\nn\\
\ee
The potential difference between $\rho=0$ and $\rho=\infty$ will diverge and hence $(P,k)$ are not sensible to be treated as thermodynamic variables when we allow $\rho$ run the full geodesic range.  On the other hand, if we restrict our boundary at $\rho=\sqrt{P/P_0}$, the thermodynamic potentials are finite, given precisely by \eqref{Phis}.

\section{The charge lattice}
\label{sec:lattice}

The BPS dyonic solution involves three parameters $(Q,P,k)$ that are subject to the constraint \eqref{gkpcons}. These parameters are associated with the electric and magnetic string charges $(Q_e,Q_m)$ and the dipole charge $Q_D$ defined by \eqref{QeQm} and \eqref{QD} respectively. In our convention \eqref{QeQm}, the electric and magnetic charges must satisfy the Dirac quantization condition \cite{Bremer:1997qb}
\be
Q_e Q_m = \frac{ n}{8}\,,\qquad n\in \mathbb{Z}\,,\label{dirac}
\ee
where Newton's constant is set to 1.
This implies that $Q_e$ and $Q_m$ should be each integer multiples of some basic unit charge.
The constraint \eqref{gkpcons} implies that $Q_D$ should be quantized as well. Since $P$ appears directly in $H_P$, it follows that it should be non-negative. It follows from \eqref{gkpcons} that we must impose
\be
0\le 2gk \le 1\,.
\ee
Applying the standard Wu-Yang argument to the 1-form gauge potential $A_{(1)}$ and using the fact fermions in the model carry charge $g$, it is reasonable to expect that the quantization of the dipole charge should be $2gk =$ integer. This suggests that the dyonic string solutions
of \cite{Gueven:2003uw} are ruled out by the Dirac quantization condition completely.

To resolve this issue of a vacuous charge lattice, we observe that the above BPS dyonic solutions is only one branch of all possible solutions. There exists another branch of dyonic strings, with the same metric and scalar functions and $(H_P, H_Q)$, but with an overall minus of the $A_\1$ and $B_\2$ fields. We may refer to this branch as dyonic anti-strings, or anti-dyons. In this case, the equations of motion require a different constraint
\be
4g P = k (1 + 2 gk)\,,\label{gkpcons2}
\ee
where positivity of $P$ is guaranteed by either $2g k>0$ or $2gk<-1$.  To see these are indeed solutions, we can start with the original string solution in $\rho$ coordinate \eqref{bpsdyonrho} and set
\be
P=-\tilde P\,,\qquad Q=-\tilde Q\,,\qquad Q_0=-\tilde Q_0\,,\qquad k=-\tilde k\,.\label{redef}
\ee
We obtain the solution in tilded variables and then drop the tilde. Thus for the same metric and scalar field, the charges now become
\be
Q_e= - \pi Q\,,\qquad Q_m= - \pi P\,,\qquad
Q_D= - \pi k\,.
\ee
The negative sign of the charges is consistent with the fact that these solutions describe anti-dyons.
It is worth pointing out that the non-extremal generalization of the anti-dyon is also straightforward. In fact, the thermodynamic quantities obtained in section \ref{sec:black} are well defined under the redefinition \eqref{redef}.

It should be pointed out that both the constraints \eqref{gkpcons} and \eqref{gkpcons2} do not involve the electric charge. This implies that the sign choice of the electric component of $B_\2$ can be both signs with the same positive $Q$ contribution to the metric and scalar. This leads to four branches of charge configurations
\bea
&&\{Q_m>0, Q_e>0, Q_D>0\}\,,\qquad \{Q_m>0, Q_e<0, Q_D>0\}\,,\nn\\
&&\{Q_m<0, Q_e<0, Q_D<0\}\,,\qquad \{Q_m<0, Q_e>0, Q_D<0\}\,.\label{4branches}
\eea
The dyonic solutions of the top line were construction in \cite{Gueven:2003uw} and the bottom line gives the new anti-dyon solutions.

We now examine the supersymmetries of these four branches of solutions. We follow same choice of the vielbein as in \cite{Gueven:2003uw} and we find that all the four branches of solutions admits Killing spinors in the form
\be
\epsilon=(H_Q H_P)^{-\fft14} \epsilon_0\,,
\ee
where the constant spinor $\epsilon_0$ are subject to two projections; therefore, the solutions all preserving $1/4$ of supersymmetry. The results are summarized in Table 1. Note that the $\pm$ signs in the table are not independent, but they are corresponds to the four-branches of solutions \eqref{4branches}.  Although these four branches of solutions has the same metric and scalar field, they have different Killing spinors, corresponding to four different ways of the projections of the constant spinor $\epsilon_0$. Therefore they cannot co-exist in the same BPS spectrum. The string/anti-string configuration will necessarily break the supersymmetry.

\bigskip
\begin{center}
\begin{tabular}{|c|c|c|}
  \hline
   & dyonic string & anit-dyonic string \\ \hline
  constraints & $4gP=k (1-2gk)$ & $4gP=k(1+2gk)$ \\ \hline
  Charges & $(Q_e,Q_m,Q_D)=\pi (\pm Q, P, k)$ & $(Q_e,Q_m,Q_D)=-\pi (\pm Q, P, k)$ \\ \hline
  Killing spinors & $(\Gamma_{1234}\pm1)\epsilon_0=0=
  (\Gamma_{12}\pm {\rm i})\epsilon_0$ &$(\Gamma_{1234}\mp 1)\epsilon_0=0=
  (\Gamma_{12}\mp {\rm i})\epsilon_0$\\ \hline
\end{tabular}

\bigskip

{\small Table 1. Constraints, charges and Killing spinors of four branches of the BPS solutions.}
\end{center}

It should be pointed out that the constraints \eqref{gkpcons} and \eqref{gkpcons2} for the dyons and anti-dyons continue to hold in the more general non-extremal black solutions. We are now in the position to derive the complete charge lattices of the dyonic and anti-dyonic strings. First, in the Mink$_4\times S^2$ vacuum, we have $2 g k=\pm 1$.  Since the $S^2$ does not shrink to zero for the general solution, we therefore have the charge quantization for the dipole parameter
\be
2g k = n\,,\qquad Q_D=\fft{\pi}{2g} n\,,\qquad n\in \mathbb{Z}\,.
\ee
This implies that there can be no dyonic string satisfying \eqref{gkpcons}. The spectrum becomes vacuous by the charge quantization. The anti-dyonic string however survives the quantization, with the magnetic charge
\be
4g^2 P = n (n+1)\,,\qquad Q_{m} = -\fft{\pi}{4g^2} n(n+1)\,,\qquad n\in \mathbb{Z}\,.
\ee
Note that $Q_m\le 0$, regardless the sign choice of $n$. Since $n(n+1)$ is always an even integer, the minimum magnetic string charge is $\pi/(2g^2)$ rather than $\pi/(4g^2)$. The charge lattice for the magnetic charges are not all negative integer multiples of this minimum charge, but instead with factors of
\be
\ft12n(n+1) = 1, 3, 6, 10, 15, 21, 28, 36, 45, 55, \cdots\,, etc.
\ee
 The Dirac quantization condition \eqref{dirac} implies that the electric charges are also
quantized, given by
\be
Q_e=\frac{g^2}{2\pi} n\,,\qquad n\in \mathbb{Z}\,.
\ee
In other words, the minimum electric charge is $g^2/(2\pi)$. Although the magnetic charges of the anti-dyon solution is always negative, the electric charge can be both positive and negative.

\section{Conclusions}
\label{sec:conclusion}

In this paper, we revisited the BPS dyonic string with magnetic dipole charge \cite{Gueven:2003uw} in Salam-Sezgin model. We made several progresses. We analyzed the global structure and showed that the geodesic is not complete with the original coordinates. We obtained the new radial coordinate that is more analogous to that of dyonic string in ungauged supergravity. We extended the BPS solutions to non-extremal black strings and obtain the first law of thermodynamics. We find that the temperature of the black strings is bounded from above, even though neither the mass nor entropy are bounded. The specific heat with fixed charge however is always positive.

We showed that the original BPS dyon of \cite{Gueven:2003uw} will not survive the Dirac quantization condition. We found that there exist a new branch of BPS solutions which we called anti-dyons that are consistent with the Dirac quantization. We therefore obtained the full charge lattice, where the magnetic string charges are all negative. One should be cautioned that the charge lattice obtained here is not directly related to those obtained in \cite{Pang:2020rir} where the lattice is determined by the anomaly coefficients. In \cite{Pang:2020rir}, the anomaly-free extensions of Salam-Sezgin model possesses only 6D, $N=(1,0)$ supersymmetry. In this framework, it could be argued that the discussion of the quantum aspects such as the charge lattice is premature for the anomalous theory \eqref{SS} when the anomaly-free theory \cite{Randjbar-Daemi:1985tdc,Avramis:2005qt} can involve higher-derivative corrections as well as the one-loop contribution to the kinetic term such as $e^{-\varphi} F_\2^2$. However, in the tree-level approximation and long-wavelength limit, the anomaly-free theory is expected to be reduced to \eqref{SS}. For our anti-dyon solutions, both extremal or non-extremal, these limits amount to setting the parameters as
\be
|Q_e Q_m|\gg 1\,,\qquad \Big|\fft{Q_e}{Q_m}\Big|\gg 1\,.\label{inequality}
\ee
(The former ensures small curvature invariants and the latter follows from \eqref{scalarsol}.) Thus the regions of our charge lattice satisfying \eqref{inequality} should survive in the extended anomaly-free theory. If Salam-Sezgin model is to be treated as a consistent truncation of certain 6D, $N=(1,1)$ theory, the issue of anomaly cancellation does not arise and the charge lattice obtained here should be considered as a sublattice of the $N=(1,1)$ theory.

As for future directions, we would like to add angular momentum into the dyonic string and analyze its effects to the thermodynamical quantities. Last but not least, the Salam-Sezgin model can be extended by curvature squared supergravity invariants \cite{Bergshoeff:2012ax, Novak:2017wqc,Butter:2018wss} and it should be interesting to see how higher derivative interactions modify properties of the dyonic string solution. A particular interest is to investigate whether the higher-derivative terms could contribute negatively to the black hole entropy, as in the case of \cite{Ma:2023qqj}.

\section*{Acknowledgement}

We are grateful to Ergin Sezgin for useful feedbacks on an earlier version of the draft. The work is supported in part by the National Natural Science Foundation of China (NSFC) grants No.~11935009 and No.~12375052 and No.~12175164. This work of Y.P. is also supported by the National Key Research and Development Program under grant No. 2022YFE0134300.


\begin{thebibliography}{99}

\bibitem{Salam:1984cj}
A.~Salam and E.~Sezgin,
``Chiral compactification on Minkowski$\times S^2$ of $N=2$ Einstein-Maxwell supergravity in six-dimensions,''
Phys. Lett. B \textbf{147}, 47 (1984)
doi:10.1016/0370-2693(84)90589-6.

\bibitem{Maeda:1984gq}
K.i.~Maeda and H.~Nishino,
``Cosmological solutions in $D=6$, $N=2$ {Kaluza-Klein} supergravity: Friedmann universe without fine tuning,''
Phys. Lett. B \textbf{154} (1985), 358-362
doi:10.1016/0370-2693(85)90409-5.

\bibitem{Maeda:1985es}
K.i.~Maeda and H.~Nishino,
``Attractor universe in six-dimensional $N=2$ supergravity {Kaluza-Klein} theory,''
Phys. Lett. B \textbf{158} (1985), 381-387
doi:10.1016/0370-2693(85)90437-X.

\bibitem{Gibbons:1986xp}
G.W.~Gibbons and P.K.~Townsend,
``Cosmological evolution of degenerate vacua,''
Nucl. Phys. B \textbf{282} (1987), 610-625
doi:10.1016/0550-3213(87)90700-0.

\bibitem{Aghababaie:2003ar}
Y.~Aghababaie, C.P.~Burgess, J.M.~Cline, H.~Firouzjahi, S.L.~Parameswaran, F.~Quevedo, G.~Tasinato and I.~Zavala,
``Warped brane worlds in six-dimensional supergravity,''
JHEP \textbf{09} (2003), 037
doi:10.1088/1126-6708/2003/09/037
[arXiv:hep-th/0308064 [hep-th]].

\bibitem{Aghababaie:2003wz}
Y.~Aghababaie, C.P.~Burgess, S.L.~Parameswaran and F.~Quevedo,
``Towards a naturally small cosmological constant from branes in 6-D supergravity,''
Nucl. Phys. B \textbf{680} (2004), 389-414
doi:10.1016/j.nuclphysb.2003.12.015
[arXiv:hep-th/0304256 [hep-th]].


\bibitem{Randjbar-Daemi:1985tdc}
S.~Randjbar-Daemi, A.~Salam, E.~Sezgin and J.A.~Strathdee,
``An anomaly free model in six-dimensions,''
Phys. Lett. B \textbf{151}, 351-356 (1985)
doi:10.1016/0370-2693(85)91653-3.

\bibitem{Avramis:2005qt}
S.D.~Avramis, A.~Kehagias and S.~Randjbar-Daemi,
``A new anomaly-free gauged super- gravity in six dimensions,''
JHEP \textbf{05}, 057 (2005)
doi:10.1088/1126-6708/2005/05/057
[arXiv:hep-th/0504033 [hep-th]].

\bibitem{Pang:2020rir}
Y.~Pang and E.~Sezgin,
``On the consistency of a class of R-symmetry gauged 6$D$ $\mathcal{N}$= (1,0) supergravities,''
Proc. Roy. Soc. Lond. A \textbf{476}, no.2240, 20200115 (2020)
doi:10.1098/ rspa.2020.0115
[arXiv:2002.04619 [hep-th]].

\bibitem{Cvetic:2003xr}
M.~Cveti\v c, G.W.~Gibbons and C.N.~Pope,
``A String and M theory origin for the Salam-Sezgin model,''
Nucl. Phys. B \textbf{677}, 164-180 (2004)
doi:10.1016/j.nuclphysb.2003.10.016
[arXiv:hep-th/0308026 [hep-th]].

\bibitem{Kerimo:2003am}
J.~Kerimo and H.~L\"u,
``New $D = 6$, ${\cal N}=(1,1)$ gauged supergravity with supersymmetric (Minkowski)$_4 \times S^2$ vacuum,'' Phys. Lett. B \textbf{576}, 219-226 (2003)
doi:10.1016/j.physletb. 2003.09.076
[arXiv:hep-th/0307222 [hep-th]].

\bibitem{Kerimo:2004md}
J.~Kerimo, J.T.~Liu, H.~L\"u and C.N.~Pope,
``Variant ${\cal N} = (1,1)$ supergravity and Minkowski$_4 \times S^2$ vacua,''
Class. Quant. Grav. \textbf{21}, 3287-3300 (2004)
doi:10.1088/0264-9381/21/13/011
[arXiv:hep-th/0401001 [hep-th]].

\bibitem{Gibbons:2003gp}
G.W.~Gibbons and C.N.~Pope,
``Consistent $S^2$ Pauli reduction of six-dimensional chiral gauged Einstein-Maxwell supergravity,''
Nucl. Phys. B \textbf{697}, 225-242 (2004)
doi:10.1016/ j.nuclphysb.2004.07.016
[arXiv:hep-th/0307052 [hep-th]].

\bibitem{Duff:1993ye}
M.J.~Duff and J.X.~Lu,
``Black and super p-branes in diverse dimensions,''
Nucl. Phys. B \textbf{416}, 301-334 (1994)
doi:10.1016/0550-3213(94)90586-X
[arXiv:hep-th/9306052 [hep-th]].

\bibitem{Duff:1996cf}
M.J.~Duff, H.~L\"u and C.N.~Pope,
``Heterotic phase transitions and singularities of the gauge dyonic string,''
Phys. Lett. B \textbf{378}, 101-106 (1996)
doi:10.1016/0370-2693(96)00420-0
[arXiv:hep-th/9603037 [hep-th]].

\bibitem{Seiberg:2011dr}
N.~Seiberg and W.~Taylor,
``Charge lattices and consistency of 6D supergravity,''
JHEP \textbf{06}, 001 (2011)
doi:10.1007/JHEP06(2011)001
[arXiv:1103.0019 [hep-th]].

\bibitem{Monnier:2017oqd}
S.~Monnier, G.W.~Moore and D.S.~Park,
``Quantization of anomaly coefficients in 6D $\mathcal{N}=(1,0)$ supergravity,''
JHEP \textbf{02}, 020 (2018)
doi:10.1007/JHEP02(2018)020
[arXiv: 1711.04777 [hep-th]].

\bibitem{Gueven:2003uw}
R.~Gueven, J.T.~Liu, C.N.~Pope and E.~Sezgin,
``Fine tuning and six-dimensional gauged $N=(1,0)$ supergravity vacua,''
Class. Quant. Grav. \textbf{21}, 1001-1014 (2004)
doi:10.1088/ 0264-9381/21/4/019
[arXiv:hep-th/0306201 [hep-th]].

\bibitem{Duff:1996hp}
M.J.~Duff, H.~L\"u and C.N.~Pope,
``The black branes of M theory,''
Phys. Lett. B \textbf{382}, 73-80 (1996)
doi:10.1016/0370-2693(96)00521-7
[arXiv:hep-th/9604052 [hep-th]].

\bibitem{Cvetic:1996gq}
M.~Cveti\v c and A.A.~Tseytlin,
``Nonextreme black holes from nonextreme intersecting M-branes,''
Nucl. Phys. B \textbf{478}, 181-198 (1996)
doi:10.1016/0550-3213(96)00411-7
[arXiv:hep-th/9606033 [hep-th]].

\bibitem{Wald:1993nt}
R.M.~Wald,
``Black hole entropy is the Noether charge,''
Phys. Rev. D \textbf{48}, no.8, R3427-R3431 (1993)
doi:10.1103/PhysRevD.48.R3427
[arXiv:gr-qc/9307038 [gr-qc]].

\bibitem{Lu:2013ura}
H.~L\"u, Y.~Pang and C.N.~Pope,
``AdS dyonic black hole and its thermodynamics,''
JHEP \textbf{11} (2013), 033
doi:10.1007/JHEP11(2013)033
[arXiv:1307.6243 [hep-th]].

\bibitem{Ma:2022nwq}
L.~Ma, Y.~Pang and H.~L\"u, ``Improved Wald formalism and first law of dyonic black strings with mixed Chern-Simons terms,''
JHEP \textbf{10} (2022), 142
doi:10.1007/JHEP10(2022)142
[arXiv:2202.08290 [hep-th]].

\bibitem{Bremer:1997qb}
M.S.~Bremer, H.~L\"u, C.N.~Pope and K.S.~Stelle,
``Dirac quantization conditions and Kaluza-Klein reduction,''
Nucl. Phys. B \textbf{529}, 259-294 (1998)
doi:10.1016/S0550-3213(98) 00369-1
[arXiv:hep-th/9710244 [hep-th]].

\bibitem{Bergshoeff:2012ax}
E.~Bergshoeff, F.~Coomans, E.~Sezgin and A.~Van Proeyen,
``Higher derivative extension of 6D chiral gauged supergravity,''
JHEP \textbf{07} (2012), 011
doi:10.1007/JHEP07(2012)011
[arXiv:1203.2975 [hep-th]].

\bibitem{Novak:2017wqc}
J.~Novak, M.~Ozkan, Y.~Pang and G.~Tartaglino-Mazzucchelli,
``Gauss-Bonnet supergravity in six dimensions,''
Phys. Rev. Lett. \textbf{119} (2017) no.11, 111602
doi:10.1103/PhysRev Lett.119.111602
[arXiv:1706.09330 [hep-th]].

\bibitem{Butter:2018wss}
D.~Butter, J.~Novak, M.~Ozkan, Y.~Pang and G.~Tartaglino-Mazzucchelli,
``Curvature squared invariants in six-dimensional ${\cal N} = (1,0)$ supergravity,''
JHEP \textbf{04} (2019), 013
doi: 10.1007/JHEP04(2019)013
[arXiv:1808.00459 [hep-th]].

\bibitem{Ma:2023qqj}
L.~Ma, Y.~Pang and H.~L\"u,
``Higher derivative contributions to black hole thermodynamics at NNLO,''
JHEP \textbf{06}, 087 (2023)
doi:10.1007/JHEP06(2023)087
[arXiv:2304.08527 [hep-th]].



\end{thebibliography}
\end{document}